\documentclass[article]{appolb}
\received{May 16, 2016; accepted September 5, 2016}

\usepackage{amsmath}
\usepackage{amssymb}
\usepackage{graphicx}

\newcommand\w{\vrule height 12 pt depth 0 pt width 0 pt}{}

\begin{document}
\title{On the D-branes Standard-like Models}
\author{{Salah Eddine:S.E.}{Ennadifi}
\address{LPHE-MS, Faculty of Science, Mohammed V University, Rabat, Morocco\\
\texttt{ennadifis@gmail.com}}}
\maketitle
\begin{abstract}
Based on the low-energy effective field theory of D-branes, the mass
spectrum of an extended Standard Model with two-Higgs doublets used to
generate all the mass terms is investigated. Besides the gauge bosons, the
fermion mass spectrum is weighted by the Higgs VEVs with a partial hierarchy
and the smallness of neutrino masses is exhibited. With reference to the
known data, the involved scales of the model are approached.
\end{abstract}

\section{Introduction}

Up to date, all low-energy phenomena have been successfully predicted by the
Standard Model of particle physics (SM) according to a vast amount of
experimental data~\cite{1,2,3,4}. The Higgs interactions with the gauge
bosons and fermions are completely determined by theory and are yet to be
experimentally established~\cite{3,4}. Despite its experimental success, the
SM is not satisfactory as a fundamental theory, and there are several
attempts with supplementary symmetries, matters or dimensions which have
been developed to address its open issues with new phenomenological
features~\cite{5,6,7}. The scalar-extended SMs constitute one of these most
well-motivated extensions bringing many interesting phenomenological features
such as fermion masses hierarchy and dark matter candidates~\cite{6,7,8,9}.

All these ingredients naturally occur in some string theory models
involving more U(1)s in the gauge group and leading to effective low-energy
theories containing more scalar fields beyond the SM. A feature of such
scalar-extended models is the existence of tree-level flavour changing
neutral currents (FCNC) whose the potentially dangerous interactions can be
avoided here by the presented extra symmetries~\cite{10,11}.

Intensive works with orientifold constructions has been recently developed
based on intersecting D-branes embedded in superstrings theory\break \cite{12,13,14,15},
where $N$ coincident D-branes typically generate a unitary group
$\mathrm{U}(N)\simeq \mathrm{SU}(N)\times \mathrm{U(1)}$ and, hence, every
stack of branes supplies the model with an extra Abelian factor in the gauge
group. Such U(1) fields have generically four-dimensional anomalies which are
cancelled via the Green--Schwarz mechanism~\cite{16,17}. This mechanism gives
a mass to the anomalous U(1) fields and breaks the associated gauge symmetry~\cite{18,19}.
In this string vacua, SM particles are considered as open string
states attached to different stacks of D-branes and their interactions are
subject to additional restriction from these remnant global Abelian U(1)
symmetries~\cite{20,21,22,23,24}. This approach provides an acceptable
effective low-energy description reproducing SM-like or some extensions and
offering a simple low $M_\mathrm{s}$ string scale energy framework.

In this paper, motivated by the progressive efforts towards obtaining the SM
in string theory constructions, we present a quiver gauge theory of a
Higgs-extended SM below a low-mass scale $M_\mathrm{s}$ based on the low-energy
effective field theory of intersecting D-branes giving rise to the gauge
symmetry $\mathrm{U}(3)\times \mathrm{SU}(2)\times \mathrm{U}(1)^{n}$.
The four-stack configuration, $n=2$, contains two Higgs doublets $H_{u}$
and $H_{d}$ to generate all fermion Yukawa couplings. Besides the gauge
bosons getting their masses from the two Higgs VEVs $v_{u}$, $v_{d}$,
the resulting fermion masses are such that the heavy fermions are weighted
by the Higgs vevs $v_{u}$, while the light fermions including neutrinos are
weighted by the second Higgs vev $v_{d}$ giving a partial solution to fermion
masses hierarchy as well as the smallness of neutrino masses. According to the
known experimental data, we bound the Higgs vevs $v_{u}$, $v_{d}$ and the
mass scale $M_\mathrm{s}$.

\section{D-brane inspiration}

\subsection{D-branes and fields content}

String theory realizations of important particle physics models have
recently grown much interest. Growing studies in gauge theories have been
developed based on intersecting D-branes embedded in Type II superstrings
where it is possible to investigate deeper details seen in particle physics\break 
\cite{13,14,15}. In these compactifications, the gauge groups
\begin{equation}
\mathrm{G}_{2+n}=\mathrm{U(3)}\times \mathrm{U(2)}\times \mathrm{U(1)}^{n}\,, \qquad
n\geqslant 0
\label{eq1}
\end{equation}
arise from stacks of D-branes that fill out four-dimensional spacetime and
wrap three-cycles in the internal Calabi--Yau threefold. Chiral matter arises
at the intersection of two different D-brane stacks in the internal space,
and their multiplicity is given by the topological intersection number of
the respective three-cycles. Their interactions are subject to additional
restrictions of the global U(1)s exhibited by orientifold
compactifications. The vacua we consider here are obtained from four stacks
of intersecting D-branes giving rise to the symmetry group\footnote{Where
the $\mbox{Sp}(1)\simeq \mbox{SU}(2)$ weak symmetry arise from
D6-wrapped on an orientifold invariant three-cycle ($b=b^{\ast }$) and
$\mathrm{U}(1)_{d}$ is a gauged flavor symmetry distinguishing various quarks
from each others.} 
\begin{equation}
\text{G}_{4}=\text{U(3)}_{a}\times \text{Sp(1)}_{b}\times \text{U(1)}_{c}\times \text{U(1)}_{d}\,.
\label{eq2}
\end{equation}
Given this brane content and conditions that we require of the SM
realizations such as tadpole cancellation and the presence of a massless
hypercharge~\cite{16,17,18,19}, we can build an intersecting brane model
with the chiral content of the SM. Since the Sp(1) does not exhibit a U(1)$_{b}$
which could contribute to the hypercharge, the mixed anomalies and
anomalous Abelian U(1)$_{a,c,d}$ parts are cancelled by the Green--Schwarz
mechanism and promoted to global U(1)$_{a,c,d}$ symmetries which are
respected by all perturbative couplings and a linear combination
U(1)$_{Y}=q_{a}$U(1)$_{a}\times q_{c}$U(1)$_{c}\times q_{d}$U(1)$_{d}$ of
them does not acquire a Stuckelberg mass and remains massless to be
identified as the hypercharge. The tadpole cancellations, which are
conditions on the cycles the D-branes wrap, imply restrictions on the
transformation properties of the chiral spectrum and guarantee the
cancellation of gauge anomalies U($N_{\alpha }$). Here, it is seen that the SU(2)
is realized as Sp(1), all representations are real and the tadpole equations
do not impose any condition on the transformation behavior under the Sp(1).
Moreover, the fact that the stack $b=b^{\ast }$ limits the potential
origins of fields charged under the $b$-brane. These conditions are used to
fit the U(1)$_{a,c,d}$ charged SM particles
\begin{eqnarray}
Q^{i} &=&\left( u,d\right)_\mathrm{L}^{i}=\left( u,d\right)_\mathrm{L},\ \left( c,s\right)_\mathrm{L},\ \left( t,b\right)_\mathrm{L}\,,\notag\\
\overline{u}^{\,i} &=&u_\mathrm{R}^{c^{i}}=\overline{u},\overline{c},\overline{t}\,,\notag\\
\overline{d}^{\,i} &=&d_\mathrm{R}^{c^{i}}=\overline{d},\overline{s}, \overline{b}\,, \notag\\
L^{i} &=&\left( v_{e},e\right) _\mathrm{L}^{i}=\left( v_{e},e\right),\ \left(v_{\mu },\mu \right),\ \left( v_{\tau },\tau \right)\,, \notag\\
\text{ }\overline{e}^{\,i} &=&e_\mathrm{R}^{c^{i}}=\overline{e},\overline{\mu}, \overline{\tau}\,.
\label{eq3}
\end{eqnarray}
In our description, the three left-handed quarks $Q^{i}$ are localized at
the intersections of D6-branes $a$ and $b$, while right-handed quarks
$\overline{u}^{\,i}$ and $\overline{d}^{\,i}$ split into two up quarks $\overline{u}^{\,2,3}$
and one down quark $\overline{d}^{\,3}$ at the intersection of the
D6-branes $a$ and $c/c^{\ast}$, and two down quarks $\overline{d}^{\,1,2}$ and
one up quark $\overline{u}^{\,1}$ at the intersection of the D6-branes $a$ and 
$d/d^{\ast}$. The three left-handed leptons $L^{i}$ arise at the
intersection of branes $b$ and $d$ respectively, and the three right-handed
electrons $\overline{e}^{\,i}$ arise at the intersection of D6-branes $d$ and
$c^{\ast}$. Finally, the two Higgs doublets $H_{c}$ and $H_{d}$ arise at
intersection of D6-branes $b$ and $c/d^{\ast}$ respectively. These
correspond to the following fermion intersection numbers\footnote{We have
not included those involving $b^{\ast}=b$. The other intersection
numbers are set to zero and as we discussed, the cancellation of $\mbox{U}(N_{a,c,d})$ anomalies~is
\begin{equation*}
\sum\limits_{\beta =a,b,c,d}N_{\beta }\left(
I_{\alpha \beta }+I_{\alpha \beta ^{\ast }}\right) =0\,.
\end{equation*}} 
\begin{eqnarray}
I_{ab} &=&3\,, \hspace{12.3mm} I_{ac}=-2\,, \qquad I_{ac^{\ast}}=-1\,, \notag\\
I_{ad} &=&-1\,, \qquad I_{ad^{\ast}}=-2\,, \hspace{9.3mm} I_{db}=3\,,  \notag\\
I_{dc^{\ast}} &=&-3
\label{eq4} 
\end{eqnarray}
which is indeed obeyed by the above spectrum. From these intersection
numbers, we summarize in the following table the fields content and the
corresponding charges which depend on the anomaly-free hypercharge linear
combination for which all the matter particles have the proper electroweak
hypercharge.

\begin{table}[htb]
{\small
\rightline{TABLE I}

\vspace{3mm}\noindent
The fields content corresponding to the anomaly-free hypercharge linear combination
$Y=\frac{1}{6}Q_{a}-\frac{1}{2}Q_{c}-\frac{1}{2}Q_{d}$. The index $i(=1,2,3)$
denotes the family index.

\vspace{3mm}\centerline{%
\begin{tabular}{l|r|r|r|r|r|r|r}
\hline\hline
\multicolumn{1}{c|}{Fields} \w &
\multicolumn{1}{c|}{$Q^{i}$}   &
\multicolumn{1}{c|}{$\overline{u}^{\,1}$}   &
\multicolumn{1}{c|}{$\overline{d}^{\,1,2}$} &
\multicolumn{1}{c|}{$\overline{u}^{\,2,3}$} &
\multicolumn{1}{c|}{$\overline{d}^{\,3}$}   &
\multicolumn{1}{c|}{$L^{i}$}                &
\multicolumn{1}{c}{$\overline{e}^{\,i}$} \\[1mm]
\hline
\w U$(1)_{c}$ & 0 & 0 & 0 & 1 & $-1$ & 0 & $-1$ \\
U$(1)_{d}$ & 0 & 1 & $-1$ & 0 & 0 & 1 & $-1$ \\
$Y$ & 1/6 & $-2/3$ & 1/3 & $-2/3$ & 1/3 & $-1/2$ & 1 \\
\end{tabular}}}
\end{table}

Since any realistic string vacua have to exhibit the phenomenologically
desired terms in its low-energy physics, we need to fit the Higgs sector.
An examination of the associated quantum numbers in Table~I shows that the
communication of the electroweak symmetry breaking to all fermion
requires, in addition to the SM Higgs doublet, a second Higgs doublet.
The latter carries the following U(1) charges, given in Table~II, 
\begin{table}[htb]\vspace{-2mm}
{\small
\rightline{TABLE II}

\vspace{3mm}\centerline{%
The Higgs sector with their U(1) charges.}

\vspace{3mm}\centerline{%
\begin{tabular}{l|r|r}
\hline\hline
\multicolumn{1}{c|}{Fields} \w & \multicolumn{1}{c|}{$H_{c}$} & \multicolumn{1}{c}{$H_{d}$} \\[1mm]
\hline
\w U$(1)_{c}$ & $-1$ & 0 \\
U$(1)_{d}$ & \ 0 & $-1$ \\
$Y$ & 1/2 & 1/2 \\
\end{tabular}}}
\end{table}
leading to one-hypercharges $Y=1/2$, two Higgs doublets $H_{c}=(H_{c}^{+},H_{c}^{0})$
and $H_{d}=(H_{d}^{+},H_{d}^{0})$, with the 99 subscripts $c$ and $d$ referring to the 
$c$-brane and $d$-brane under which the 100 Higgses are charged.

\begin{figure}[htb]
\centerline{%
\includegraphics[width=8.3cm]{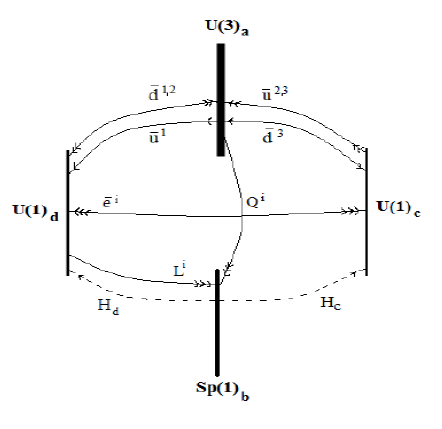}}
\caption{Four-stack quiver SM.}
\label{fig1}
\end{figure}

Based on these data and string theory description, this model can be
obtained from an orientifold compactification with D-branes wrapping
non-trivial cycles as represented in Fig.~\ref{fig1}
where bold lines represent the relevant D-branes, which are not
distinguished from their orientifold images, giving rise to the gauge
symmetry of~Eq.~(\ref{eq2}); solid and dashed thin lines indicate the
chiral intersections of those branes and refer to matter fields and Higgs
fields respectively; arrows directions indicate fundamental
(antifundamental) representations of the U$(N)$ gauge groups.

Although this is a simple scalar extension of the SM with two
one-hypercharge Higgses and thus with more parameters than in the 
minimal SM, it provides a rich mass spectrum.

\section{Mass spectrum}

\subsection{Higgs spectrum and electroweak boson masses}

The Higgs scalar fields of the model consist of two complex isodoublets, or
eight real, scalar degrees of freedom. We have then in terms of physical
fields
\begin{equation}
H_{i=c,d}=\left( 
\begin{array}{c}
H_{i}^{+} \\ 
H_{i}^{0}
\end{array}
\right) =\left( 
\begin{array}{c}
h_{1i}^{+}+ih_{2i}^{+} \\ [1mm]
\left( v_{i}+h_{1i}^{0}+ih_{2i}^{0}\right)
\end{array}
\right)\,.
\label{eq5}
\end{equation}
When the electroweak symmetry is broken by the VEVs of the two Higgses
$v_{c} $ and $v_{d}$ minimizing the corresponding renormalizable and gauge
invariant Higgs potential
\begin{eqnarray}
V\left( H_{c},H_{d}\right) &=&\mu _{c}^{2}\left( H_{c}^{\dagger
}H_{c}\right) +\mu _{d}^{2}\left( H_{d}^{\dagger }H_{d}\right) +\half
\lambda _{c}\left( H_{c}^{\dagger }H_{c}\right) ^{2}+\half\lambda
_{d}\left( H_{d}^{\dagger }H_{d}\right) ^{2}  \notag \\
&&+\lambda _{cd}\left( H_{c}^{\dagger }H_{c}\right) \left( H_{d}^{\dagger
}H_{d}\right) +\lambda _{cd}^{\prime }\left( H_{c}^{\dagger }H_{d}\right)
\left( H_{d}^{\dagger }H_{c}\right)\,,
\label{eq6}
\end{eqnarray}
three of the eight real scalar degrees are the would-be charged and neutral
Nambu--Goldstone bosons $G^{\pm }$ (mixture of $h_{1i}^{+}$ and
$h_{2i}^{+})$ and $G^{0}$ (mixture of $h_{2i}^{0}$) which become the
longitudinal modes of the $Z_{\mu }$ $=g_{2}W_{\mu }^{3}-gB_{\mu }/
\sqrt{g_{2}^{2}+g_{1}^{1}}$ and $W_{\mu }^{\pm }=W_{\mu }^{1}\mp iW_{\mu }^{2}/
\sqrt{2}$ electroweak bosons getting masses through the kinetic part of the
Higgs potential
\begin{eqnarray}
&&\left \vert D_{\mu }H_{c}\right \vert ^{2}+\left \vert D_{\mu }H_{d}\right
\vert ^{2} =\left \vert \left( \partial _{\mu }+ig_{2}\frac{\tau _{a}}{2}
W_{\mu }^{a}-ig_{1}\frac{1}{2}B_{\mu }\right) \right \vert ^{2}\left( \left
\vert H_{c}\right \vert ^{2}+\left \vert H_{d}\right \vert ^{2}\right) 
\notag \\
&&=\frac{1}{4}g_{2}^{2}\left( v_{c}^{2}+v_{d}^{2}\right) ^{2}W_{\mu
}^{+}W^{-\mu }+\frac{1}{8}\left( g_{2}^{2}+g_{1}^{2}\right) \left(
v_{c}^{2}+v_{d}^{2}\right) ^{2}Z_{\mu }Z^{\mu }+ \dots
\label{eq7}
\end{eqnarray}
With the SM constraint $\sqrt{v_{c}^{2}+v_{d}^{2}}=v=\left( G_\mathrm{F}\sqrt{2}
\,\right) ^{-1/2}=246$~GeV limiting the parameters of the Higgs potential, the
electroweak boson masses are
\begin{equation}
m_{W}=\half gv\,,\qquad m_{Z}=\half\sqrt{g_{2}^{2}+g_{1}^{2}}v\,.
\label{eq8}
\end{equation}
The remaining five degrees of freedom are the physical Higgs bosons; they
consist of three neutral scalars $h_{1}^{0}$, $h_{1}^{0^{\prime }}$, 
$h_{2}^{0}$ (mixture of $h_{1i}^{0}$, $h_{2i}^{0}$ with $m_{h_{2}^{0}},$\break
$m_{h_{1}^{0^{\prime }}} > m_{h_{1}^{0}}$) and two charged scalars 
$h^{\pm }$ (mixture of $h_{1i}^{+},$ $h_{2i}^{+}$ with $m_{h_{1}^{0^{\prime
}}}>m_{h_{1}^{0}}$) which are important from an experimental
point of view.

\subsection{Fermion masses and hierarchy}

In all 2HDMs, the most serious potential problem faced is the possibility of
FCNC which can cause severe phenomenological difficulties~\cite{10,11}.
Nonetheless, under reasonable assumptions, models with these problems may
still be viable. In the presented model, the tree-level FCNC are completely
absent due to the extra continuous symmetries U(1)$_{c,d}$ distinguishing
the SM fermions such that all fermions with the same quantum numbers couple
to the same Higgs multiplet~\cite{10,11}. Since the presence of the two
Higgs doublets was required to generate the Yukawa couplings for the all
fermions, according to the D-brane model building selection rules, the
allowed quark and lepton Yukawas are
\begin{eqnarray}
Q_{c,d} \!\!&=&\!\!\left(H_{c}Q^{i}\overline{u}^{\,2,3}\right)\!=\!Q_{c,d}
\left(\!H_{c}^{\dagger }Q^{i}\overline{d}^{\,3}\right)\!=\!Q_{c,d}\left( H_{d}Q^{i}
\overline{u}^{\,1}\right)\!=\!Q_{c,d}\left(\!H_{d}^{\dagger }Q^{i}\overline{d}^{\,1,2}\right)\!\!=\!0\,, \notag\\
Q_{c,d}\!\!&=&\!\!M_\mathrm{s}^{-1}\left(H_{d}L^{i}\right)^{2}\!=\!Q_{c,d}\left( H_{c}^{\dagger}L^{i}\overline{e}^{\,i}\right)\!=\!0\,,
\label{eq9}
\end{eqnarray}
and thus the corresponding Yukawa Lagrangian for the three families of
quarks and leptons including left-handed neutrinos reads
\begin{eqnarray}
L_\mathrm{Y} &=&y_{u^{2,3}}H_{c}Q^{i}\overline{u}^{\,2,3}+y_{d^{3}}H_{c}^{\dagger
}Q^{i}\overline{d}^{\,3}+\text{\ }y_{e^{i}}H_{c}^{\dagger }L^{i}\overline{e}
^{\,i}+y_{u^{1}}H_{d}Q^{i}\overline{u}^{\,1}  \notag\\
&&+y_{d^{1,2}}H_{d}^{\dagger}Q^{i} \overline{d}^{\,1,2} +y_{\nu ^{i}}M_\mathrm{s}^{-1}\left(H_{d}L^{i}\right)^{2}\,,
\label{eq10}
\end{eqnarray}
where $y$s are coupling constants and $M_\mathrm{s}$ being the string
mass scale. After the electroweak symmetry breaking and explicating the
resulting coupling terms by using the\ attributed fermion family indices,
we get
\begin{eqnarray}
L_\mathrm{Y_{mass}} &=&\left(y_{c}Q\overline{c}+y_{t}Q\overline{t}+y_{b}Q
\overline{b}+y_{e^{i}}L^{i}\overline{e}^{\,i}\right) v_{c}+\left( y_{u}Q
\overline{u}+y_{d}Q\overline{d}+y_{s}Q\overline{s}\right) v_{d}  \notag \\
&&+\left( \frac{y_{\nu ^{i}}}{M_\mathrm{s}}L^{i}L^{i}\right) v_{d}^{2}\,.
\label{eq11}
\end{eqnarray}
This Lagrangian exhibits clearly 3 fermion mass scales. The first related to~$v_{c}$,
the second to $v_{d}$ and the third to the highly suppressed term
$v_{d}^{2}/M_\mathrm{s}$. The resulting fermion masses read then
\begin{eqnarray}
m_{c} &=&y_{c}v_{c}\,, \hspace{10mm} m_{b}=y_{b}v_{c}\,, \hspace{9.5mm} m_{\tau }=y_{t}v_{c}\,, \notag\\
m_{u} &=&y_{u}v_{d}\,, \hspace{9.3mm} m_{d}=y_{d}v_{d}\,, \hspace{9mm} m_{s}=y_{s}v_{d}\,, \notag\\
m_{e^{i}} &=&y_{e^{i}}v_{c}\,, \qquad m_{v^{i}}=\frac{1}{M_\mathrm{s}} y_{v^{i}}v_{d}^{2}\,.
\label{eq12}
\end{eqnarray}
In this view, with the Yukawa constants hierarchy: $y_{t}\sim 1>y_{c}>y_{b}>y_{s}\gg y_{d}\sim y_{u}\gg y_{e^{i}}$
and the extremely small effective neutrino Yukawa constant $y_{v^{i}}v_{d}/M_\mathrm{s}\ll y_{e^{i}}$,
the heavy known quarks $t$, $b$ and $c$
and leptons $e^{i}$ $=$ $e$, $\mu,\tau$ get their masses from the first
Higgs VEV $v_{c}$, the light known quarks $u$, $d$ and $s$ get their masses
from the second Higgs VEV $v_{d}$ and the left-handed neutrinos
$v^{i}=v^{e},v^{\mu }$, $v^{\tau }$ get their masses from the second Higgs
VEV $v_{d}$ through a high-dimension operator suppressed by the mass scale
$M_\mathrm{s}$ explaining their tiny masses. This leads to
\begin{equation}
v_{c}\sim \mathrm{GeV} > v_{d}\sim \mathrm{MeV} \gg \frac{v_{d}^{2}}{M_\mathrm{s}}\sim \mathrm{eV}
\label{eq13}
\end{equation}
which shows that we have indeed three scales responsible for the generation
of fermion masses and giving a partial explanation for the observed mass
hierarchies. Since the mass scale $M_\mathrm{s}$ is taken as the low-string scale
at which neutrino masses have origin, it is interesting to approximate its
value. For neutrino masses upper bound $m_{\nu_{\tau}}\lesssim 2$~eV from~(\ref{eq12}), we can get
\begin{equation}
M_\mathrm{s}\sim 10^{6}~\mathrm{GeV}\,.
\label{eq14}
\end{equation}
This is a stringy prediction for new physics beyond SM that we have been
able to derive from the known data; it would be useful to ask what is the
physical implication of these deductions for low-energy physics and whether
the Large Hadron Collider experiment is able to confirm the stringy physics
directly.

\section{Conclusions}

In this work, we have discussed a string-inspired extended SM where the
corresponding effective low-energy theory emerges from intersecting
\mbox{D-branes}. The interaction terms are subject to additional restrictions from
U(1) symmetries and anomaly cancellation conditions.\ In particular, we have
discussed a four stacks of intersecting D-brane configuration in
orientifolded geometries realizing the SM spectrum with two Higgs doublets
necessary for generating the Yukawa couplings for all fermion families.
The extended Higgs sector generates rich mass spectrum and presents interesting
features. In addition to the charged physical Higgses that are
experimentally important, the fermion masses have been split into three
scales GeV, MeV and eV, that we have bounded from the known data.
This hierarchy offers a partial solution to the fermion masses problem as well as the
smallness of neutrino masses.

This is among others one simple string-inspired model for flavours that is
able to address some open issues in the SM within the allowed window by
assuming that the scale of new physics, taken as the string scale, is
closely related to the scale at which the neutrino masses are generated.

\vspace{7mm}
The author would like to thank URAC09/CNRST for support.

\flushleft

\end{document}